\documentclass[epj]{svjour}
\usepackage{graphicx}
\begin{document}
\title{Superconductivity of SrTiO$_{3-\delta}$}

\author{M. Jourdan \and N. Bl\"umer \and H. Adrian}
\offprints{jourdan@uni-mainz.de}          % Insert a name or remove this line
\institute{Institute of Physics, Johannes Gutenberg University,
55099 Mainz, Germany}
\date{Received 4 December 2002 / Received in final form 10 March 2003\\
Published in Eur.\ Phys.\ J.\ B {\bf 33}, 25 (2003)}
\abstract{Superconducting SrTiO$_{3-\delta}$ was obtained by
annealing single crystalline SrTiO$_3$ samples in ultra high
vacuum. An analysis of the $V(I)$ characteristics revealed very
small critical currents $I_{\mathrm{c}}$ which can be traced back
to an unavoidable doping inhomogeneity. $R(T)$ curves were
measured for a range of magnetic fields $B$ at $I\ll
I_{\mathrm{c}}$, thereby probing only the sample regions with the
highest doping level. The resulting curves $B_{\mathrm{c2}}(T)$
show upward curvature, both at small and strong doping. These
results are discussed in the context of bipolaronic and
conventional superconductivity with Fermi surface anisotropy. We
conclude that the special superconducting properties of
SrTiO$_{3-\delta}$ can be related to its Fermi surface and compare
this finding with properties of the recently discovered
superconductor MgB$_2$.
\PACS{
      {74.70.Dd}{Superconducting materials, ternary compounds}   \and
      {74.20.Mn}{Nonconventional mechanisms}   \and
      {74.25.Fy}{Transport properties}
     } % end of PACS codes
} %end of abstract
\maketitle
\section{Introduction}
It is well known that doped SrTiO$_3$ becomes superconducting with
a transition temperature $T_c$ which strongly depends on the
doping level \cite{Sch65}. Early theories of superconductivity in
materials with small charge carrier densities considered a
many-valley semiconductor band structure as beneficial for
superconductivity due to an increased electron-phonon scattering
rate \cite {Par69}. Indeed, first band structure calculations
\cite{Kah64} predicted a Fermi surface of six disjoint ellipsoids
for doped SrTiO$_3$. However, later calculations \cite{Mat72},
which are in good agreement with experimental results
\cite{Gre79}, exhibited a Fermi surface of two sheets at the zone
center.
% revised:
Despite of the low charge carrier density of doped SrTiO$_3$ and
assuming no extraordinary electron-phonon scattering rate the
appearance of superconductivity can be explained by a two-phonon
mechanism \cite{Nga74}. Alternatively, a recent theoretical
treatment predicts strong electron-phonon coupling due to reduced
electronic screening at low doping \cite{Jar00}.\\
% end revised
Due to the specific band structure and reduced interband
scattering, so called two band superconductivity is possible in
SrTiO$_3$. This expression describes the formation of distinct
superconducting order parameters on the two sheets of the Fermi
surface. Evidence for two band superconductivity in Nb-doped
SrTiO$_3$, but not in SrTiO$_{3-\delta}$, was found by tunneling
spectroscopy performed with an STM \cite{Bin80}. Recently two band
superconductivity attracted considerable interest because it was
proposed to be realized in the new
superconductor MgB$_2$ \cite{Gui01}.\\
Independently of the band structure considerations given above,
doped SrTiO$_3$ is regarded as a candidate material for
bipolaronic superconductivity \cite{Mic90}. The basic idea is that
two polarons form a local pair called bipolaron. These bipolarons
are charged bosons which can condense into a superfluid-like
state. The possibility of bipolaronic superconductivity was
extensively discussed for high temperature superconductors, but
presumably cannot be applied to these materials \cite{Cha98}.
However, in doped SrTiO$_3$ due to its huge dielectric
polarizability ($\epsilon \simeq 300$) and low carrier density
($\simeq 10^{20}$cm$^{-1}$) \cite{Sch65} the possibility of
polaron formation is obvious and experimental evidence for their
formation exists \cite{Ger93,Eag96,Ang00}. However, pairing of
polarons and the formation of itinerant bipolarons was not
observed in any compound up to now.\\
In this paper we first describe the preparation of superconducting
SrTiO$_{3-\delta}$. It is shown that the doping of our samples is
inhomogeneous. However, from $V(I)$ curves we determine a current
which is small enough to probe only the superconducting surface
region. Thus it is possible to measure the temperature dependence
of the upper critical field of SrTiO$_{3-\delta}$. $B_{c2}(T)$
curves of different samples are compared with the theories of
bipolaron superconductivity and conventional superconductivity
including Fermi surface anisotropy.

\section{Preparation}
\begin{figure}%[htb]
\centerline{
\includegraphics[width=0.8\columnwidth,clip=true]{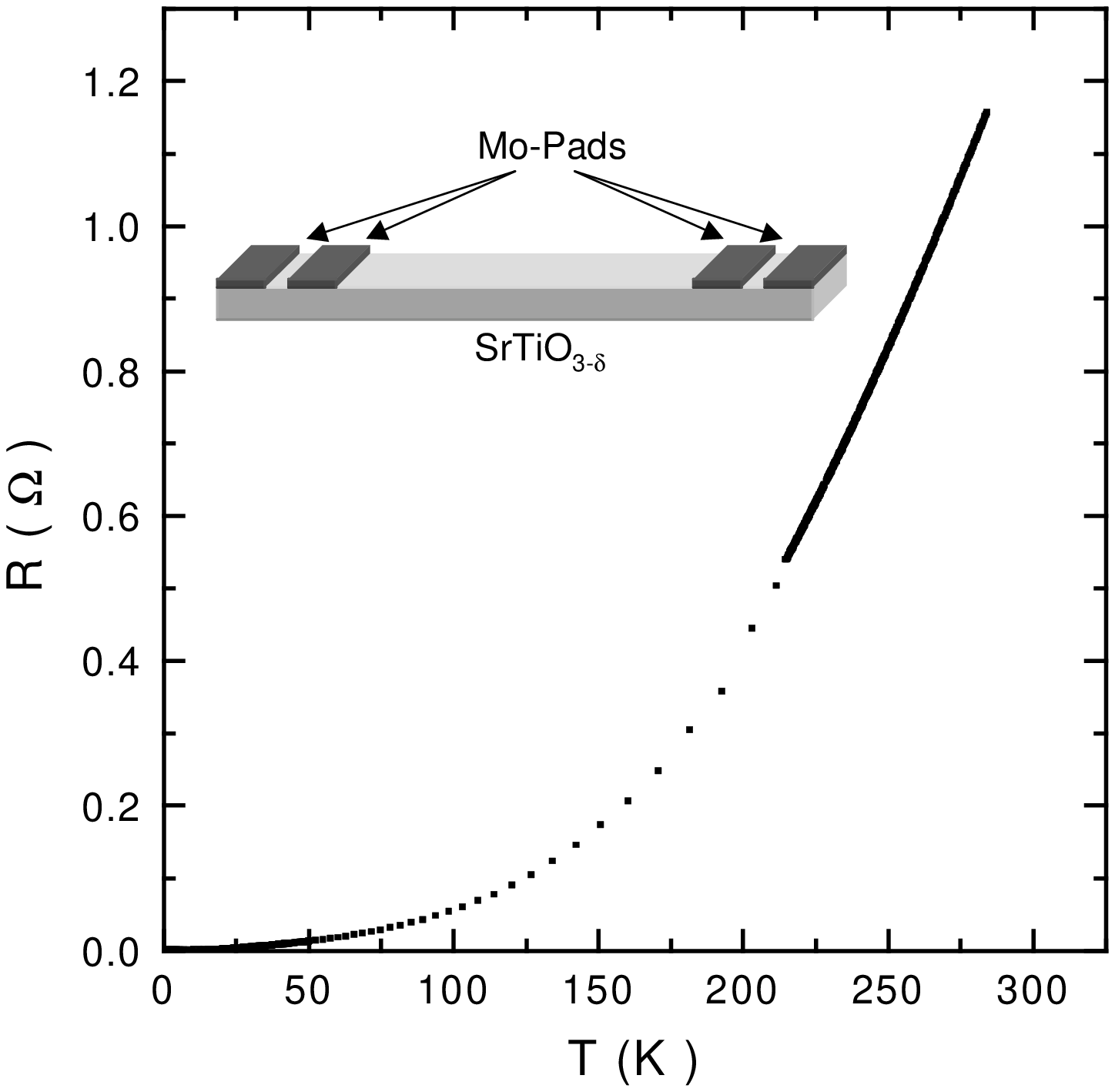}}
\caption{\label{RTuSchema} Temperature dependence of the
resistance of metallic SrTiO$_{3-\delta}$. Inset: Schematics of a
SrTiO$_{3-\delta}$ sample with contact pads (not to scale). }
\end{figure}

The starting point of our preparation of SrTiO$_{3-\delta}$
samples were commercially available single crystalline SrTiO$_3$
(111) substrates of size $10\times10\times1$mm.
%revised
The special substrate orientation was chosen because the
generation of oxygen vacancies by annealing of (111) oriented
SrTiO$_3$ proved to be more effective than that of (001) oriented
samples.
%end revised
To obtain a geometry suitable for resistance measurements the
substrates were polished down to a thickness of $\simeq 300{\rm
\mu m}$ and cut into bars of width $\simeq 700{\rm \mu m}$. Two
small Mo contact pads were deposited in a MBE chamber on each end
of the SrTiO$_3$ stripes to allow 4-probe measurements of the
sample resistance (see inset of Fig.\,\ref{RTuSchema}).
The process of oxygen reduction was performed in a MBE chamber
($p\simeq10^{-8}$mbar) at
temperatures $T\simeq1150$K for 1h.\\
Varying the annealing temperature by the amount of $\Delta T
\simeq 50$K it was possible to tune the properties of the samples
continuously from semiconducting to metallic \cite{Jou02}. A
typical temperature dependence of the resistance of a metallic
sample is shown in Fig.\,\ref{RTuSchema}. All samples which showed
metallic properties down to lowest temperatures became
superconducting and $T_c$ decreased with decreasing annealing
temperature.
% revised
Employing the results of ref.\ \cite{Sch65} from the obtained
$T_c$ values the charge carrier density of our samples can by
estimated to be between $10^{18}$cm$^{-3}$ and $10^{19}$cm$^{-3}$.
This is equivalent to a number of oxygen deficiencies per site of
$\delta \simeq$$10^{-5}$ - $10^{-4}$. Because of this very small
number of vacancies their clustering is not likely to appear and a
rigid band shift of the Fermi level into the conduction band is
expected \cite{Sha98}.
% end revised
The highest critical temperature obtained was $T_c=230$ mK. The
excellent purity of our samples is reflected by residual
resistance ratios of up to $RRR=R_{300K}/R_{0.3K}=3000$. If a
homogeneous current distribution in the sample is assumed,
specific residual resistances of $\rho(0.3{\rm K}) \simeq 1 \times
10^{-4}{\rm  \Omega cm}$ are obtained.\\
However, recently it was pointed out that the doping level of
reduced SrTiO$_{3-\delta}$ is inhomogeneous \cite{Szo02}. Reducing
the thickness of an annealed sample by mechanical polishing could
show that its specific room temperature resistance was
macroscopically constant over its volume. Nevertheless, we
observed that superconductivity is limited to a surface layer.
When $\simeq 50{\rm \mu m}$ of SrTiO$_{3-\delta}$ were removed
from the backside (side closest to the heating element) of an
annealed sample a strong reduction of $T_c$ was obtained. Removing
an additional $\simeq 50{\rm \mu m}$ layer from the front side
destroyed superconductivity completely. However, removing only
$\simeq 0.3 {\rm \mu m}$ from both sides of an annealed sample by
ion beam etching does not change its superconducting critical
temperature. Thus we conclude that the oxygen content of our
SrTiO$_{3-\delta}$ bars does not vary as strongly as in the
samples of ref.\,\cite{Szo02}. Still, the doping level close to
the surface is increased compared to the bulk volume.

\section{V(I)-curves}
\begin{figure}%[htb]
\centerline{
\includegraphics[width=0.85\columnwidth]{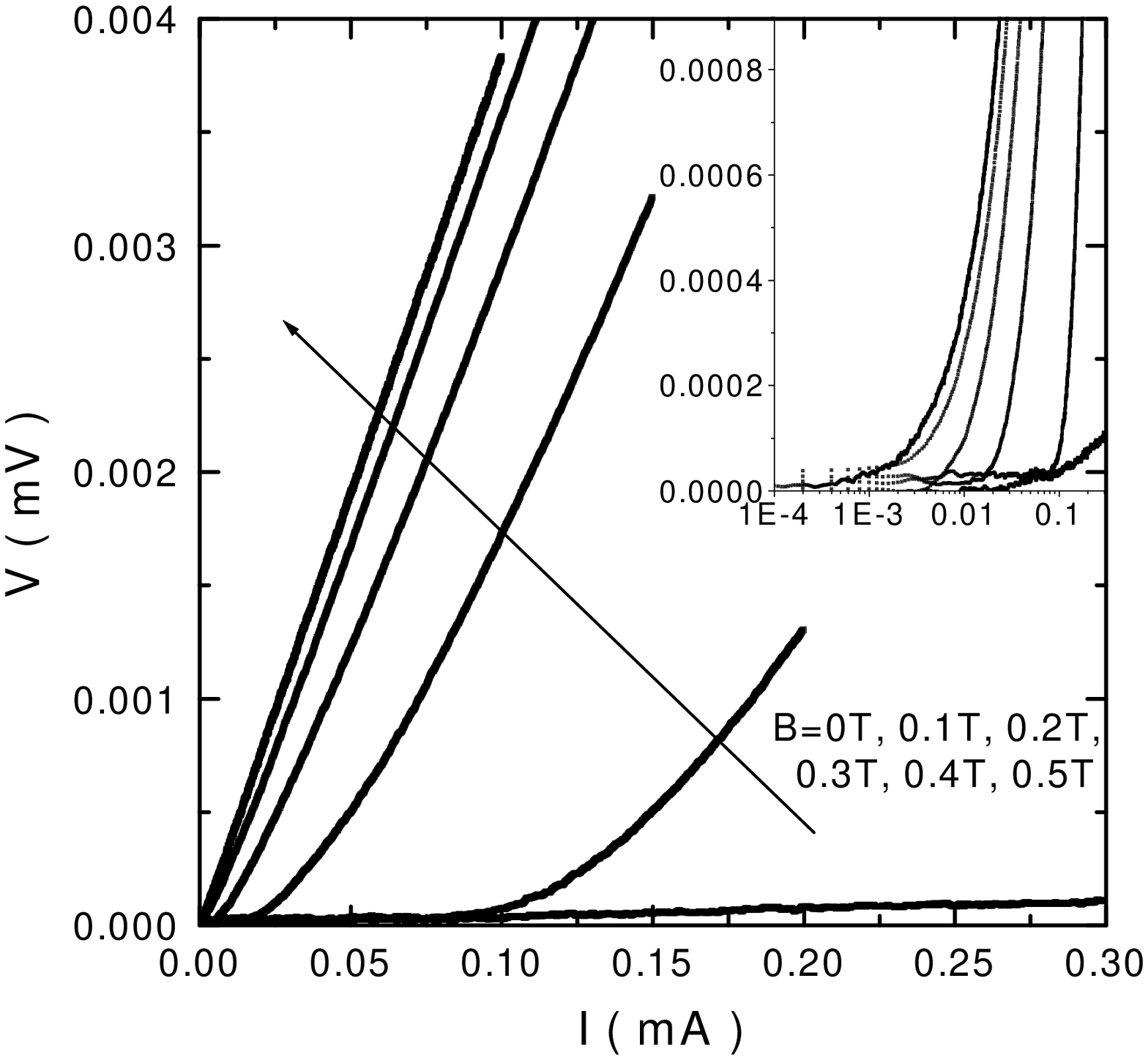}}
\caption{\label{VImeas} $V(I)$-curves of superconducting
SrTiO$_{3-\delta}$ in different applied magnetic fields $B$.}
\end{figure}

\begin{figure}%[hbt]
\centerline{
\includegraphics[width=0.8\columnwidth,clip=true]{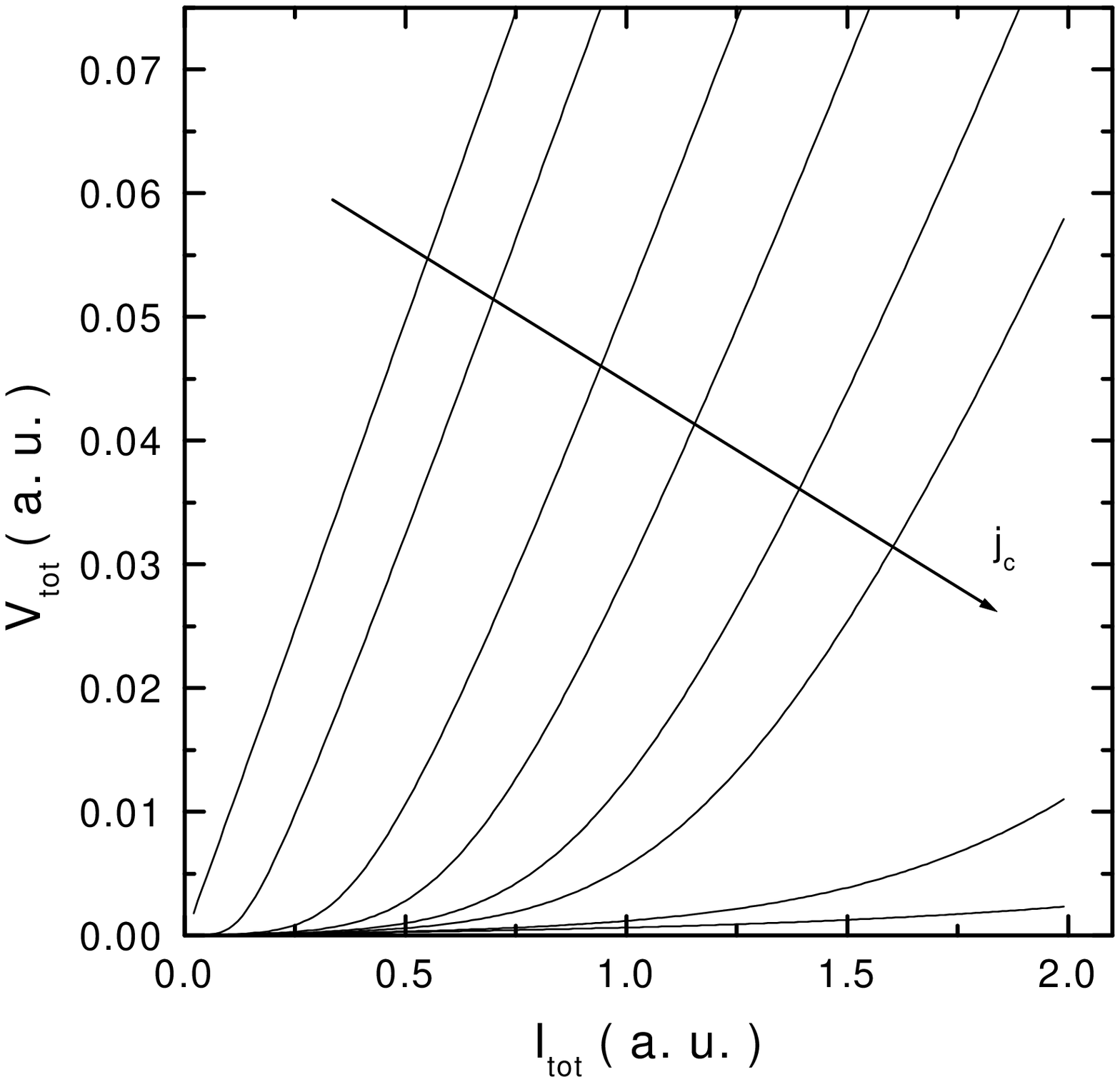}}
\caption{\label{VIcalc} Calculated $V(I)$-curves of a parallel
circuit of a normal conductor and a superconductor with different
critical current densities $j_c$ .}
\end{figure}

For further investigation of the homogeneity of the
superconducting state an increasing current was sent through the
samples and the resulting voltage was measured. The obtained
V(I)-curves are shown in Fig.\,\ref{VImeas}. If a homogeneous
current distribution in the SrTiO$_{3-\delta}$ bars is assumed,
the calculated critical current densities will become very small
and will be only of the order of $j_c \simeq 100{\rm A/cm^2}$.
However, from the measured $V(I)$-curves it can be concluded that
our samples consist of parallel normal conducting regions and
superconducting regions. Thus the superconducting path for $T<T_c$
always remains in the critical regime and a current density and
magnetic field dependent part of the total current flows in the
normal conducting path. Qualitatively the measured $V(I)$-curves
can be reproduced by a simple model: The sample is described by a
parallel ohmic resistance $R_N$ and a type-II superconductor with
a $V(I)$-characteristics governed by flux line dynamics. One of
the simplest models for the description of these dynamics given by
Anderson and Kim \cite{And64} is based on thermally activated
motion of pinned flux lines. According to this description the
electric field $E$ in the superconductor is given by equation \ref{VI}
with the current density $j$, critical current density $j_c$,
pinning potential $U_P$, specific
normal resistance $\rho_c$  and temperature T.
\begin{equation}
\label{VI}
 E(j)=2 \rho_c j_c e^{-U_P/k_BT}\sinh{\frac{jU_P}{j_ck_BT}}\,.
\end{equation}
Fig.\,\ref{VIcalc} shows calculated $V_{tot}(I_{tot})$-curves of a
parallel normal conductor of resistance $R_N$ and a superconductor
of the resistance given by equation \ref{VI}.
The qualitative agreement with the measured curves of
Fig.\,\ref{VImeas} is obvious. Since neither the specific
resistances of the normal and superconducting path nor the pinning
potential of the flux lines is known, quantitative fits of the
data cannot be performed.\\
However, the observed doping inhomogeneity has important
implications for the measurement of the upper critical field as
will be shown in the next section.

\section{Upper critical field $\bf B_{c2}(T)$}
\begin{figure}%[htb]
\centerline{
\includegraphics[width=0.82\columnwidth,clip=true]{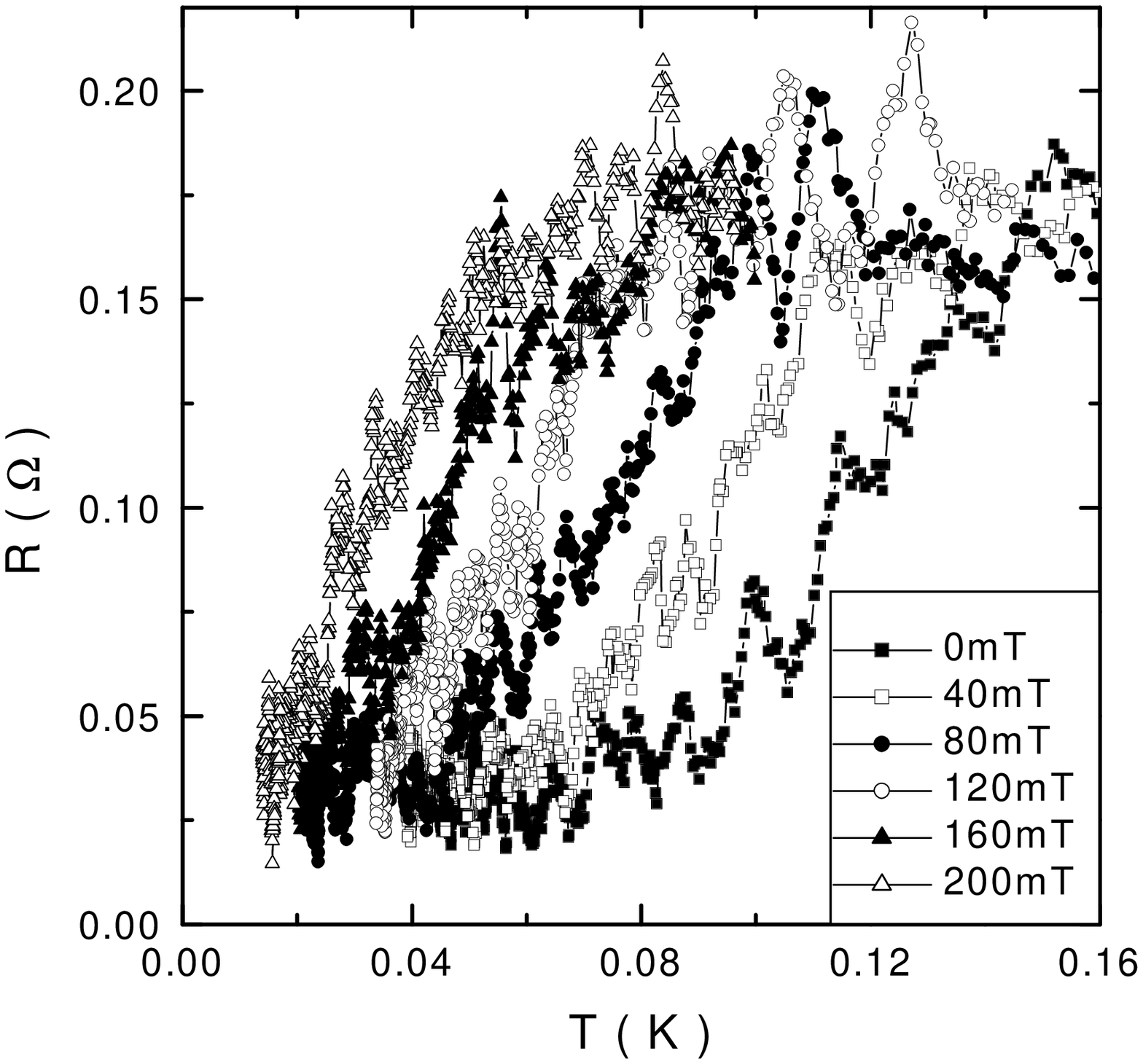}}
\caption{\label{SC_RT_LT} $R(T)$-curves of superconducting
SrTiO$_{3-\delta}$ with low doping level in various applied
magnetic fields $B$. Current perpendicular to magnetic field,
magnetic field parallel (111).}
\end{figure}

\begin{figure}[htb]
\centerline{
\includegraphics[width=0.79\columnwidth,clip=true]{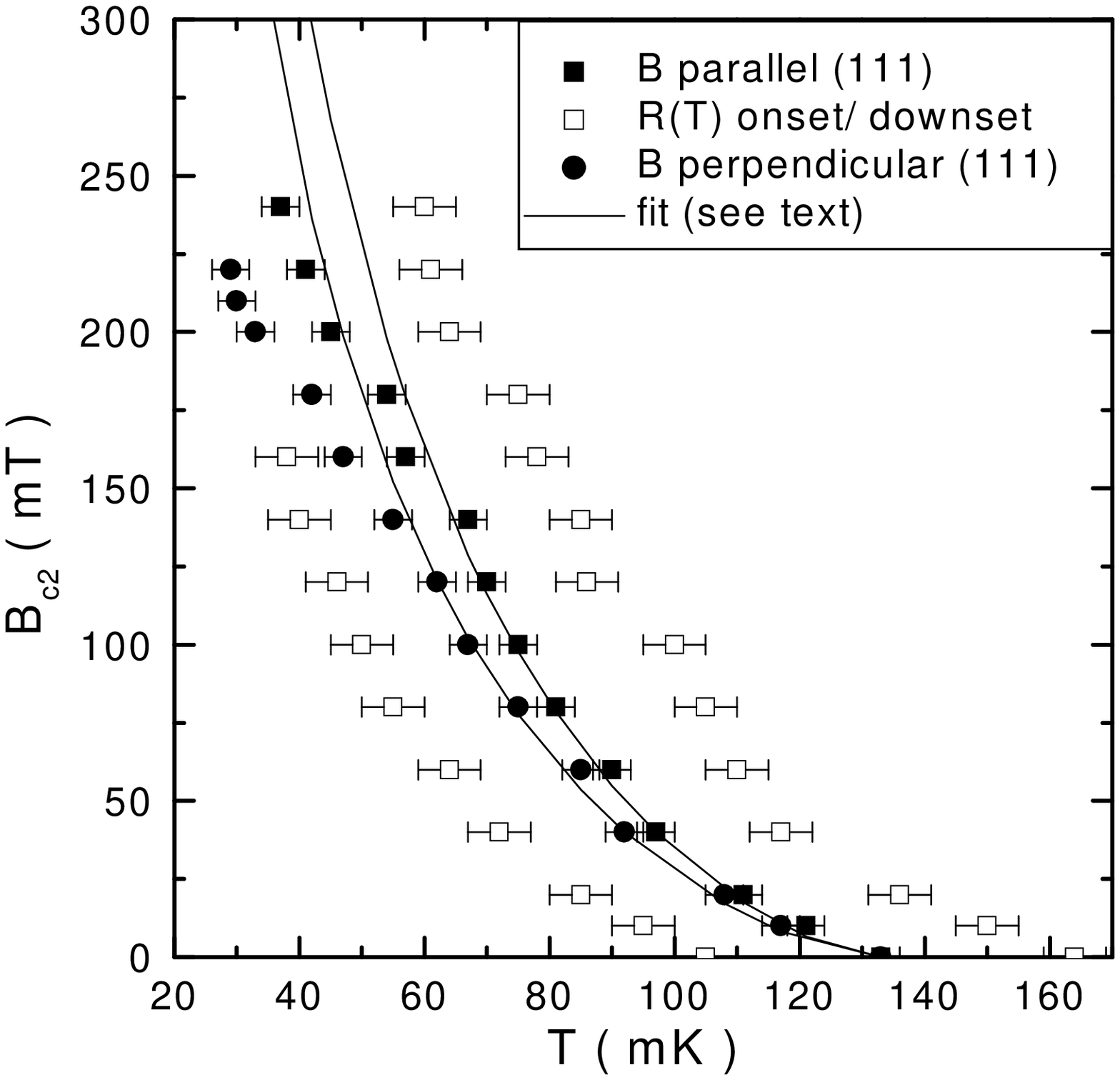}}
\caption{\label{BT_LT} $B_{c2}(T)$-curves of superconducting
SrTiO$_{3-\delta}$ with low doping level. Current direction
perpendicular to the magnetic field. Field orientation parallel to
sample surface, {\it i.e.}\ perpendicular (111), or perpendicular to
sample surface, {\it i.e.}\ parallel (111). The fits shown are
according to a theory \cite{Ale87} of bipolaronic
superconductivity (see text).}
\end{figure}

\begin{figure}%[htb]
\centerline{
\includegraphics[width=0.8\columnwidth,clip=true]{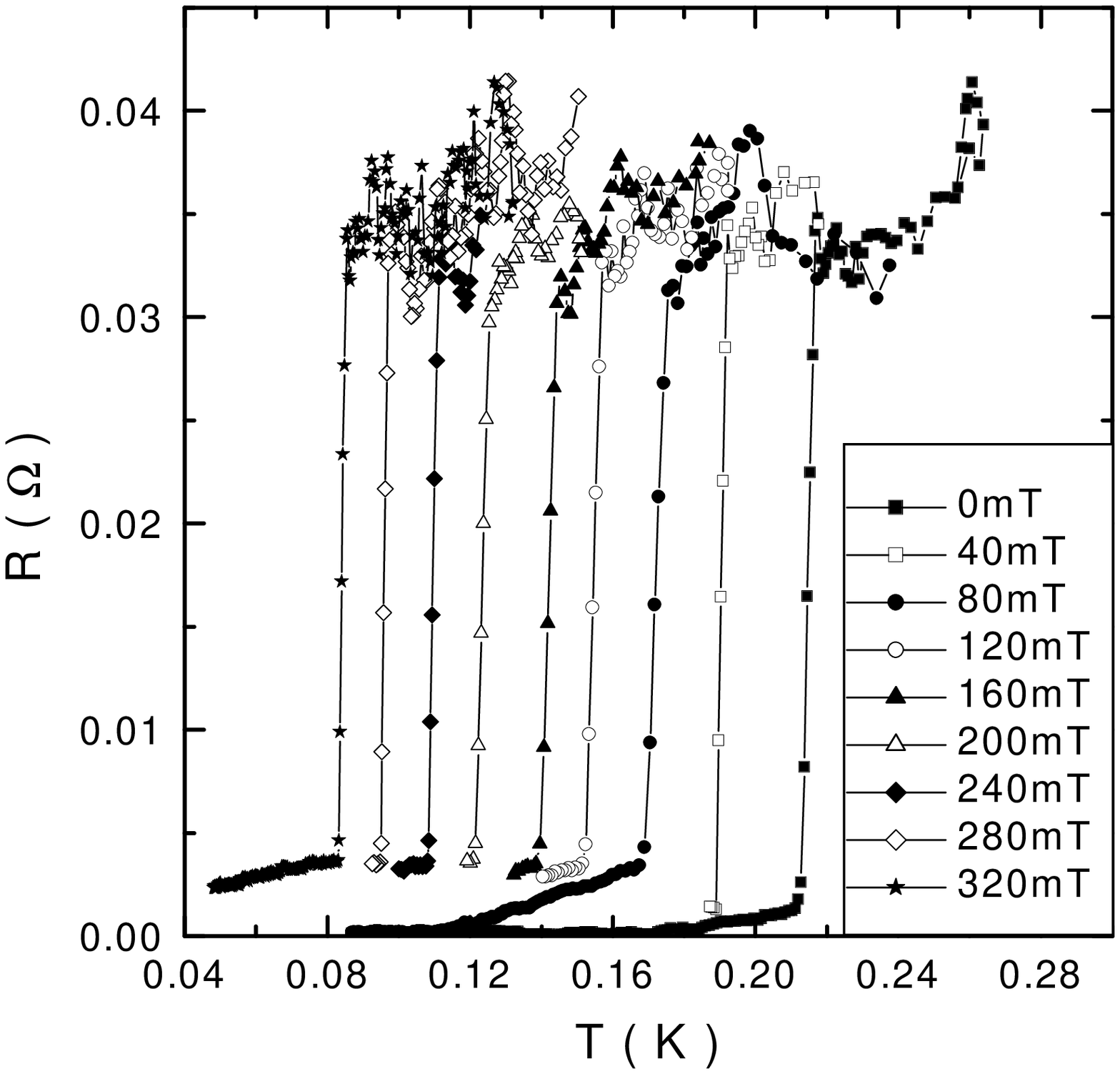}}
\caption{\label{SC_RT_HT} $R(T)$-curves of superconducting
SrTiO$_{3-\delta}$ with high doping level in various applied
magnetic fields $B$. Current perpendicular to magnetic field,
magnetic field parallel (111).}
\end{figure}

\begin{figure}%[htb]
\phantom{x}
\centerline{
\includegraphics[width=0.82\columnwidth]{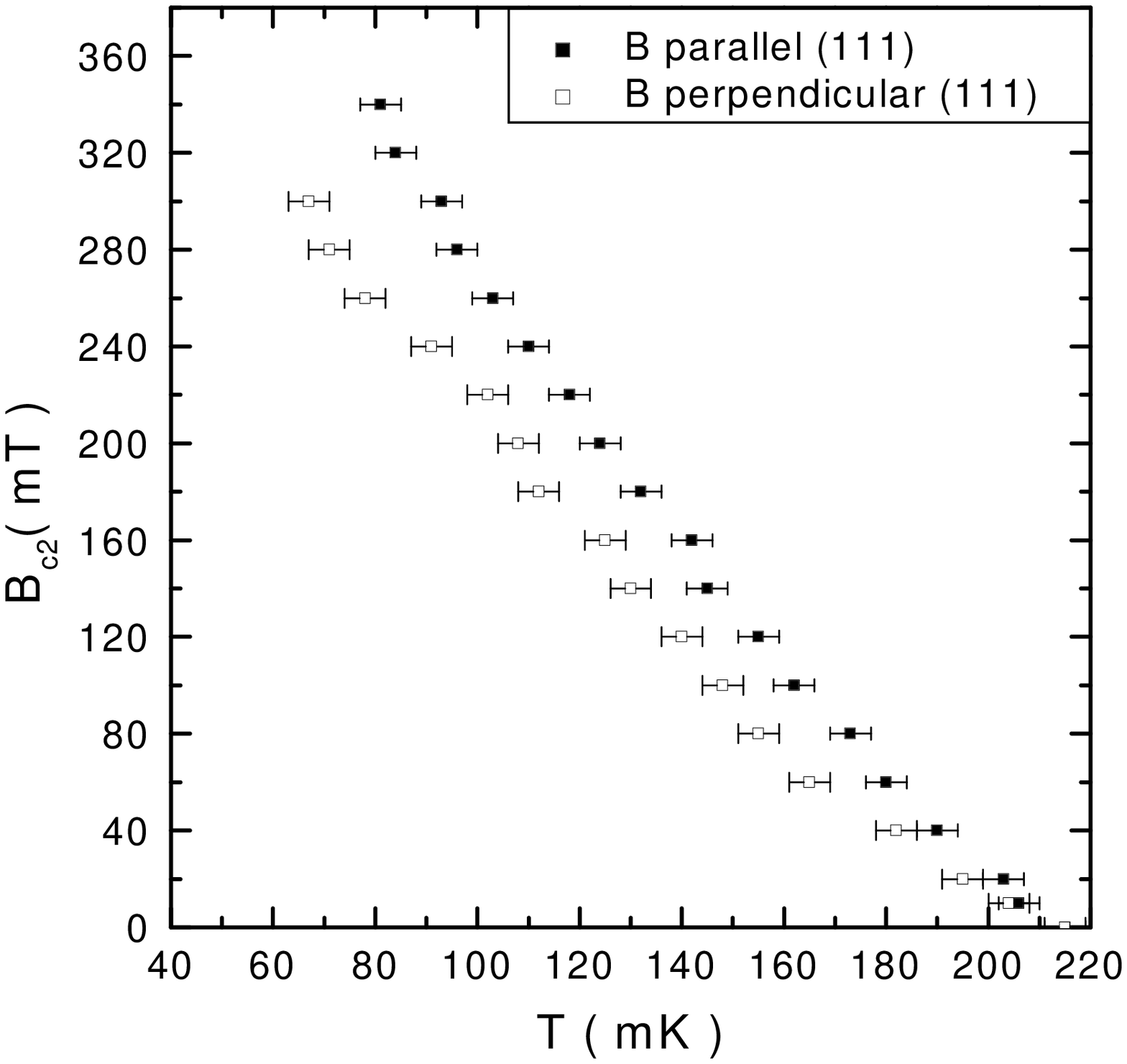}}
\caption{\label{BT_HT} $B_{c2}(T)$-curves of superconducting
SrTiO$_{3-\delta}$ with high doping level. Current direction
perpendicular to the magnetic field. Field orientation parallel to
sample surface, {\it i.e.}\ perpendicular (111), or perpendicular to
sample surface, {\it i.e.}\ parallel (111).}
\end{figure}

\begin{figure}%[htb]
\centerline{
\includegraphics[width=0.82\columnwidth,clip=true]{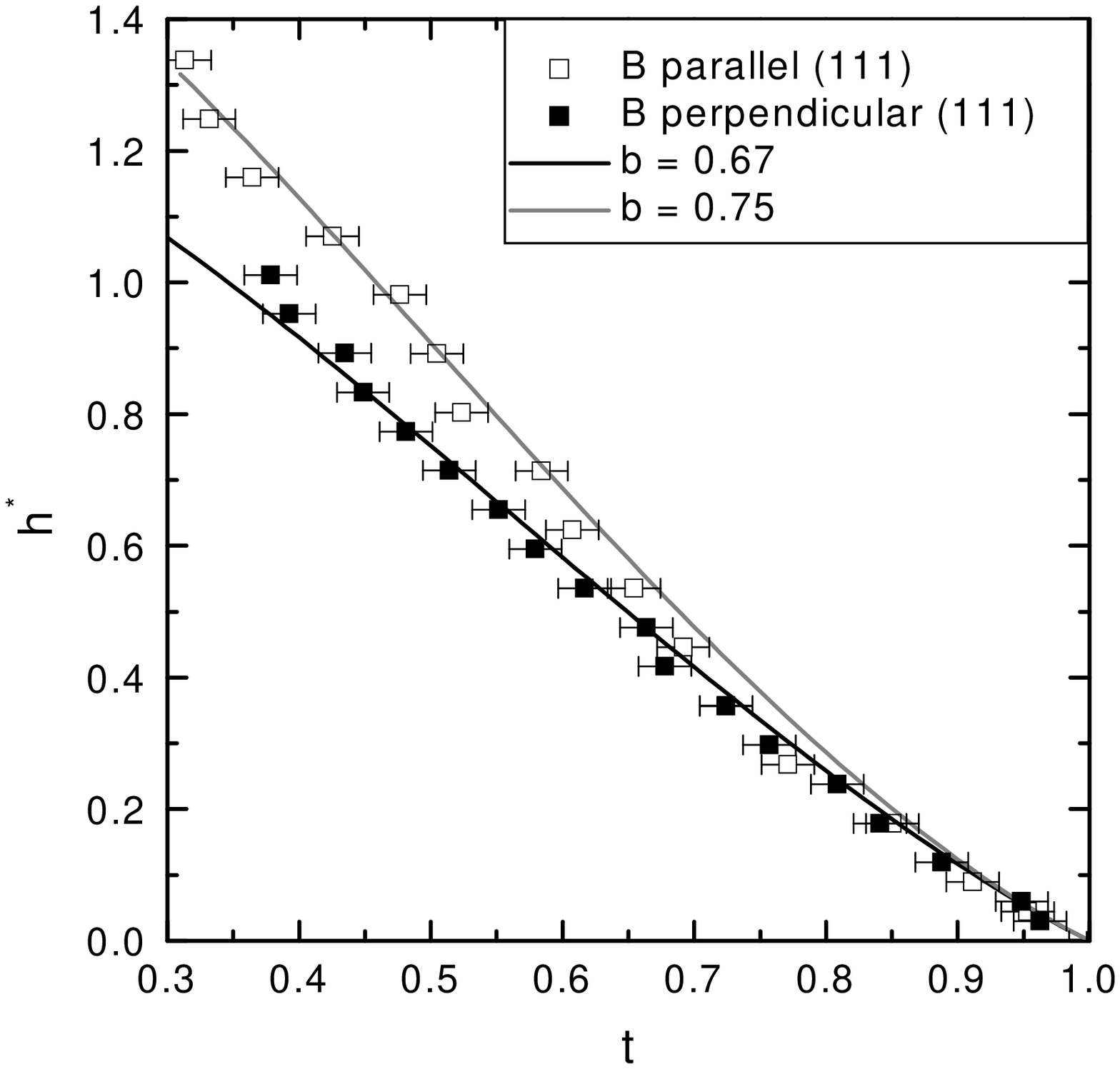}}
\caption{\label{fig8} Comparison between the normalized
temperature dependence of the upper critical field $h^*(t)$ and
fits based on the linearized Gor'kov gap equation with different
anisotropy parameter $b$.}
\end{figure}

From measurements of the upper critical field $B_{c2}(T)$
conclusions concerning Fermi surface and order parameter symmetry
or even the mechanism of superconductivity are possible.
Considering the inhomogeneities of the doping level of
SrTiO$_{3-\delta}$ these measurements have to be performed
carefully. From the analysis given above it is known that the
samples consist of parallel normal conducting and superconducting
regions with a distribution of critical temperatures $T_c$. Thus
when $T_c(B)$ is determined by measuring resistance curves $R(T)$
or $R(B)$ it is important that the probe current does not produce
local critical current effects. Provided a small enough current,
the resistance of the current path with the highest doping level,
{\it i.e.}\ in our range of doping
highest superconducting $T_c$, dominates the result.\\
$B_{c2}(T)$ curves were determined by measuring the temperature
dependence of the sample resistivity $R(T)$ in various magnetic
fields. The probe current was chosen to be always  more than an
order of magnitude below the critical current $I_c$ as estimated
from the strong increase in the $V(I)$-curve shown in the inset of
Fig.\,\ref{VImeas}. Additionally it was checked that temperature
and width of the superconducting transition did not change when
the probe current was doubled. The measured $R(T)$ curves of
samples with low doping level show a small residual resistance in
the superconducting state. This observation has to be related to
the inhomogeneous doping (see above) and the fact that the contact
wires were bonded to the upper sample surface. With a low doping
level only the lower sample surface is superconducting and the
resistance of the bulk material between the contact pads and lower
sample surface persists in the superconducting state.\\
Samples with low $T_c$ in zero field ($T_c \simeq 100$ mK) showed
broad superconducting transitions (Fig.\,\ref{SC_RT_LT}).
The width of these transition slightly shrinks from $\Delta T
\simeq 45$ mK to $\Delta T \simeq 35$ mK when the magnetic field is
increased. $B_{c2}(T)$ can be determined either by choosing a
midpoint, onset or downset criterion of the $R(T)$ curves.
However, the choice of this criterion does only shift the upper
critical field curve on the temperature axis but does not alter
the qualitative features of its temperature dependence.
% revised
In Fig.\,\ref{BT_LT} $B_{c2}(T)$ curves are shown, which are
determined by an onset (90\%), midpoint (50\%), and downset (10\%)
criterion. (B perpendicular (111): Only midpoint criterion shown
for clarity).
% end revised
This SrTiO$_{3-\delta}$ sample has a low doping level and the
magnetic field orientation was parallel or perpendicular to the
sample surface, {\it i.e.}\ perpendicular or parallel to the
crystallographic (111)-direction.
The observed directional anisotropy may either reflect
crystallographic anisotropy (cubic to tetragonal phase transition
at $T=110K$ \cite{Mat72}) or the depth profile of the doping
level. However, the fact that the critical field oriented
perpendicular to the polished sample surface is larger than
oriented parallel to the sample proves that nucleation of
superconductivity at the sample surface ($B_{c3}$) has no
influence and that actually $B_{c2}$ is measured.\\
The most striking feature of the $B_{c2}(T)$ curves of samples
with low doping level is their pronounced positive curvature. Such
a positive curvature of the temperature dependence of the upper
critical field is predicted by a theory of bipolaronic
superconductivity. According to Alexandrov et al. \cite{Ale87}
$B_{c2}(T)$ should be given by equation \ref{Ale}
\begin{equation}
\label{Ale} B_{c2}(T)\propto \frac{1}{T} \left(
1-\left(\frac{T}{T_c}\right)^{3/2} \right)^{3/2}\,.
\end{equation}
The divergence at $T=0$K in this equation results from an
approximation in ref.\,\cite{Ale87}. Fits according to
equation \ref{Ale} are plotted in Fig.\,\ref{BT_LT}. Reasonable
agreement with the measured data is observed for magnetic fields
$B<150$mT. For increased magnetic fields the fit deviates from the
measured data which could be explained by the approximations of
theory mentioned above. However, alternative explanations for the
observed positive
curvature in $B_{c2}(T)$ are possible as will be shown below.\\
More reliable $B_{c2}(T)$ data could be obtained for
SrTiO$_{3-\delta}$ samples with increased doping level due to a
strongly reduced and magnetic field independent superconducting
transition width of $\Delta T \simeq 3$ mK. Measured $R(T)$ curves
in various magnetic fields of a sample with high doping level are
shown in Fig.\,\ref{SC_RT_HT}.
The foot of the resistive transition has to be related to the
inhomogeneous doping as well. Obviously, the backside of the
sample has an increased doping level, whereas the front side with
the contact pads shows a reduced $T_c$ with broad transition
(lower doping level). Based on the $R(T)$ data the temperature
dependence of the upper critical field is determined by the
midpoints of the sharp superconducting transitions
(Fig.\,\ref{BT_HT}).
Only a weak positive curvature of $B_{c2}(T)$ is obtained, much
less pronounced than from samples of reduced doping level. Thus it
is not possible to fit this data with the bipolaron theory
(Eq.\,\ref{Ale}).\\
% revised
% Since the mechanism of superconductivity in SrTiO$_{3-\delta}$ is
% assumed to be independent of the doping level, the applicability
% of the bipolaron theory is improbable for
% samples with low doping level as well.\\
% end revised
\sloppy
The conventional theory of the upper critical field in isotropic
weak coupling superconductors by Helfand and Werthammer
\cite{Hel66} predicts a negative curvature of $B_{c2}(T)$ for all
temperatures. However, this also is in contradiction to the
$B_{c2}(T)$ data obtained for SrTiO$_{3-\delta}$. For an
applicable theoretical description in the framework of
conventional superconductivity the strongly anisotropic Fermi
surface of
SrTiO$_{3-\delta}$ has to be taken into account.\\
For simplicity we consider only orbital pair breaking and a single
sheet anisotropic Fermi surface. A formulation for this case on
the basis of the linearized Gor'kov gap equation has been put
forward by Youngner and Klemm \cite{You80}. In the clean limit, it
reduces to the self-consistency equation
\begin{equation}\label{eq:consist}
\ln t = \sum_{\nu=-\infty}^\infty
\left[\frac{1}{|2\nu+1|}\left(-1+\int d\hat{\rho}
N(\hat{\rho})f(z_{\nu,\hat{\rho}})\right)\right]\,,
\end{equation}
where $t=T_{\mathrm{c}}/T_{\mathrm{c0}}$ is the dimensionless
critical temperature,
\begin{equation}\label{eq:fz}
f(z_{\nu,\hat{\rho}})=\sqrt{\pi}z_{\nu,\hat{\rho}}
\exp(z_{\nu,\hat{\rho}}^2) \,{\rm erfc}(z_{\nu,\hat{\rho}})\,,
\end{equation}
and
\begin{equation}
z_{\nu,\hat{\rho}}=\frac{t\,|2\nu+1|}{\sqrt{2h}} \frac{v_{{\rm
F}}}{|v_\perp(\hat{\rho})|}\,.
\end{equation}
Here, the dimensionless magnetic field
\begin{equation}\label{eq:h}
h=H_{{\rm c2}}\,e\hbar v_{{\rm F}}^2/(2\pi k_{{\rm B}} T_{{\rm
c0}})^2
\end{equation}
is defined in terms of some average Fermi velocity $v_{{\rm F}}$
and the zero-field critical temperature; its absolute scale will
be unimportant for the final results. Furthermore,
$v_\perp(\hat{\rho})$ denotes the component of the Fermi velocity
in direction $\hat{\rho}$ which is perpendicular to the magnetic
field. A simple model for an anisotropic Fermi surface with
tetragonal symmetry (with the magnetic field in $(001)$ direction)
is given in polar coordinates by
\begin{equation}
|v_\perp(\hat{\rho})|=v_F \sin(\Theta) [1+b \cos(4\phi)]\,,\quad
\end{equation}
\begin{equation}
N(\hat{\rho})\propto \frac{1}{1+b \cos(4\phi)}\,,
\end{equation}
where $0\le b<1$ is the anisotropy parameter and the density of
states is normalized to unity: $\int d\hat{\rho} N(\hat{\rho})=1$.
We solve Eq.~\ref{eq:consist} for $h$ at fixed $t$ in a bracketing
secant scheme until the error in $t$ ({\it i.e.}, the difference between
the exponentiated right hand side of Eq.~\ref{eq:consist} and $t$)
is smaller than $10^{-4}$. Within this scheme, we use the cutoff $
|\nu| \le 300 $; the irreducible part of the Fermi surface ($0 \le
\cos\Theta \le 1$, $0 \le \phi \le \pi/4$) is sampled by $10^3$
patches. The maximum relative error in $h$ introduced by cutoff
and discretization is estimated as $10^{-3}$. In order to avoid
numerical instabilities, we use in equation \ref{eq:fz} the asymptotic
expansion $f(z) \approx 1 -\frac{1}{2}z^{-2} + \frac{3}{4}z^{-4}$
for $z>20$. A comparison between theory and experiment is shown in
Fig.\,\ref{fig8}.
Here, the experimental data has been converted to a dimensionless
temperature using $T_{{\rm c0}}=214$ mK. Furthermore, all magnetic
fields have been rescaled so that the derivative
$dh^*/dt|_{t=1}=-1$ for each curve; note that this rescaling
procedure is somewhat ambiguous for the experimental data.
Excellent agreement is observed between the data measured parallel
to $(111)$ and the theory for a value $b=0.67$ of the anisotropy
parameter.  This value implies a ratio of 5 of the maximum to the
minimum Fermi velocity which is well within the limits discussed
in the literature \cite{Mat72}.  The fact that the theory for
magnetic field in $(001)$ direction agrees so well with the
experimental observation for field in $(111)$ direction supports
our assumption that the overall variation of the Fermi velocity
(and of the density of states) is more important than the precise
shape of its distribution.  Less satisfactory agreement is seen
when the magnetic field is in-plane, {\it i.e.}, perpendicular to the
$(111)$ direction. Here, a stronger anisotropy $b=0.75$ is
required for matching the region $t < 0.6$; larger discrepancies
remain near $t \simeq 0.75$. An application of this theory to the
$B_{c2}(T)$ data of SrTiO$_{3-\delta}$ samples with low doping
level fails to reproduce the measured strong curvature in the same
temperature range.\\
% revised
In principle the qualitative difference of the $B_{c2}(T)$ data of
samples with low and high doping could be associated with a
crossover of the mechanism of superconductivity. For bipolaronic
superconductivity above some critical doping it is generally
expected that superconductivity is destroyed \cite{Ale94}.
However, the doping level of our samples is well below the doping
level with maximum $T_c$ \cite{Sch65}. Thus if bipolaronic
superconductivity is realized in doped SrTiO$_3$, we would expect
it to be realized in all of our samples. Assuming a doping
independent mechanism of superconductivity in SrTiO$_{3-\delta}$,
the bipolaron theory can not be applied at all.\\
One explanation for the discrepancies between samples with low and
high doping is inhomogeneous doping of the samples with low charge
carrier density resulting in domains with different $T_c$, which
influence the curvature of $B_{c2}(T)$. However, the samples with
increased charge carrier density show sharp superconducting
transitions, {\it i.e.}\ a homogeneously doped sample region is
probed. The $B_{c2}(T)$ curves of these samples could be fitted by
a conventional theory including Fermi-surface anisotropy. Thus it
appears likely that the mechanism of superconductivity of
SrTiO$_{3-\delta}$
is conventional independent of the doping level.\\
The observed positive curvature of $B_{c2}(T)$ is not unique to
SrTiO$_{3-\delta}$. An alternative approach results in very
similar curves and considers an effective two-band model, {\it i.e.}\
two groups of electrons with different superconducting energy gaps
\cite{Shu98} (applied for borocarbides). Similar behavior was
observed in the recently discovered superconductor MgB$_2$
\cite{Lim01,Sol02}. There is strong evidence for the contribution
of two different areas of the Fermi surface to the superconducting
state of this compound \cite{Iav02,Bou02}. Very recently,
$B_{c2}(T)$ curves of MgB$_2$ with positive curvature were
calculated considering superconducting gaps on two Fermi sheets as
a particular case of gap anisotropy \cite{Mir03}.
% end revised

\section{Conclusions}
\vspace{-0.5ex}
Considering the properties of self doped, {\it i.e.}\ ultra high
vacuum annealed, SrTiO$_{3-\delta}$ it is necessary to take an
inhomogeneous doping profile into account. Nevertheless, by
careful adjustment of the probe current the upper critical field
of high purity samples could be investigated. Samples with low
doping level show a strong positive curvature of the
$B_{c2}(T)$-curve, which is still present, but less pronounced for
samples with increased doping level. The theory of bipolaronic
superconductivity predicts a positive curvature of $B_{c2}(T)$ and
can be used for a fit of reasonable agreement with the
$B_{c2}(T)$-curves of samples with low doping level.
%revised
However, more reliable data obtained from samples with increased
doping level can not be fitted by the bipolaron theory, which
discards its applicability.
%end revised
A conventional explanation for the observed temperature dependence
of the upper critical field can be given by considering the
strongly anisotropic Fermi surface of SrTiO$_{3-\delta}$. Good
fits of $B_{c2}(T)$ were obtained using a concept which goes
beyond the Helfand-Werthamer theory by taking this Fermi surface
anisotropy into account.
% revised
Thus the unusual superconducting properties of SrTiO$_{3-\delta}$
can be explained in the framework of a conventional pairing
mechanism.
% end revised

\vspace{-0.5ex}
%
% BibTeX users please use
% \bibliographystyle{}
% \bibliography{}

\end{document}